\documentclass[letter]{aa}  
\usepackage{graphicx}
\usepackage{amsmath}
\usepackage{amsfonts}
\usepackage{amssymb}
\usepackage{gensymb}
\usepackage{textcomp}
\usepackage[varg]{txfonts}
\usepackage{natbib}
\usepackage{epstopdf}
\usepackage[colorlinks,linkcolor=black,linktocpage=true,plainpages=false,
urlcolor=black,citecolor=black,hypertexnames=false,breaklinks]{hyperref}
\bibpunct{(}{)}{;}{a}{}{,}
\begin{document}                               


\title{Hot prominence spicules launched from turbulent cool solar prominences}

\author{L.~P.~Chitta\inst{1}, H.~Peter\inst{1}, \and L.~Li\inst{2,3}}

   \institute{Max-Planck-Institut f\"ur Sonnensystemforschung, Justus-von-Liebig-Weg 3, 37077 G\"ottingen, Germany\\
              \email{chitta@mps.mpg.de} 
              \and
              CAS Key Laboratory of Solar Activity, National Astronomical Observatories, Chinese Academy of Sciences, Beijing 100101, People’s Republic of China
              \and
              University of Chinese Academy of Sciences, Beijing 100049, People’s Republic of China
              }

   \date{Received 5 June 2019 / Accepted 18 June 2019}

\abstract
{A solar filament is a dense cool condensation that is supported and thermally insulated by magnetic fields in the rarefied hot corona. Its evolution and stability, leading to either an eruption or disappearance, depend on its coupling with the surrounding hot corona through a thin transition region, where the temperature steeply rises. However, the heating and dynamics of this transition region remain elusive. We report extreme-ultraviolet observations of quiescent filaments from the Solar Dynamics Observatory that reveal prominence spicules propagating through the transition region of the filament-corona system. These thin needle-like jet features are generated and heated to at least 0.7\,MK by turbulent motions of the material in the filament. We suggest that the prominence spicules continuously channel the heated mass into the corona and aid in the filament evaporation and decay. Our results shed light on the turbulence-driven heating in magnetized condensations that are commonly observed on the Sun and in the interstellar medium.}

   \titlerunning{Hot prominence spicules launched from turbulent cool solar prominences}
   \authorrunning{L. P. Chitta et al.}

   \maketitle

\section{Introduction} \label{sec:intro}

Magnetized condensations of turbulent plasma in the form of elongated filaments are common in astrophysics. Their occurrence ranges from the vast interstellar medium \citep{2010A&A...518L.102A} to the confines of the solar atmosphere \citep{2014LRSP...11....1P}, where their dynamics can be investigated in greater detail. Only on the Sun can we cover their full life cycle. A solar condensation comprising cooler $10^4$\,K plasma is supported and thermally insulated by helical magnetic fields in the hotter $10^6$\,K corona \citep{1989ApJ...343..971V,2010SSRv..151..333M,2018LRSP...15....7G}. 

The condensed filament and the corona are coupled through a thin transition region where the temperature steeply rises by two orders of magnitude. Plasma dynamics and heating of this transition region are thought to maintain conditions for global and local equilibria that govern the long-term evolution and stability of a filament or its off-limb counterpart, the prominence. This is particularly the case for quiescent condensations confined by weaker magnetic fields that evolve on timescales of days to weeks \citep{2010SSRv..151..243L,2015ASSL..415.....V}. The steep temperature rise at the prominence-corona interface\footnote{Hereafter, we use the terms interface and transition region, interchangeably.} is similar to the case of the well-known gravitationally stratified chromosphere-corona transition region. Mass and energy exchange between the latter are regulated by magnetoconvective heating from the solar photosphere. This heating is marked by signatures such as chromospheric spicules \citep{1968SoPh....3..367B,2007PASJ...59S.655D} and jets \citep{2014Sci...346A.315T} that channel magnetohydrodynamic (MHD) waves \citep{2011Natur.475..477M} and quasi-periodic intensity disturbances that propagate from the chromosphere into the corona \citep{2015ApJ...815L..16S}. 

\begin{figure*}
\begin{center}
\includegraphics[width=120mm]{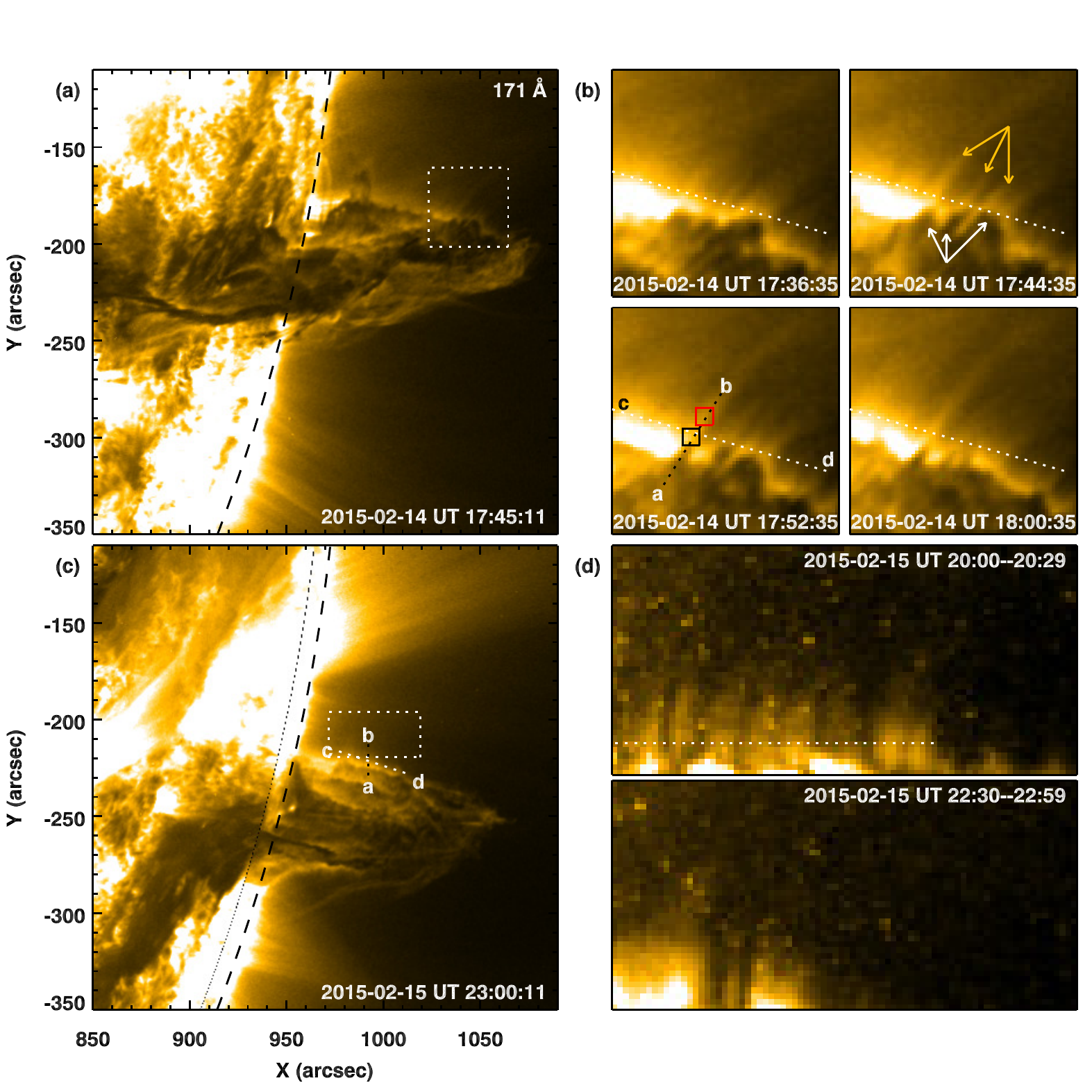}
\caption{Imaging observations of prominence spicules from a solar prominence. \textit{Panel a}: Coronal EUV emission map recorded by the SDO/AIA 171\,\AA\ showing plasma at temperatures of 0.7\,MK and more. The map covers a large prominence, seen as an absorption feature, suspended in the corona over the southwest limb of the Sun on 2015-02-14. The dotted box (field of view of 41\arcsec$\times$41\arcsec) identifies a portion of the prominence-corona transition region displayed in \textit{panel b}. The solar limb is identified with the black dashed curve. 
\textit{Panel b}: Time sequence of turbulent motions (white upward-pointing arrows) and prominence spicules (yellow downward-pointing arrows) at the prominence-corona transition region. The dynamics along a prominence spicule identified with slit a-b and the two regions (marked with black and red boxes) are shown in Fig.\,\ref{fig:pd1}. Slit c-d is overlaid as a reference to visualize the out- and inward turbulent motions of the prominence. Wave motions of prominence spicules crossing this slit are shown in Fig.\,\ref{fig:wave1}. 
\textit{Panel c}: Same as \textit{panel a,} but plotted for the observations obtained on 2015 February 15. The dotted box (field of view of 48\arcsec$\times$24\arcsec) identifies a portion of the prominence-corona transition region displayed in \textit{panel d} (the dotted curve is an artifact created by missing data in these pixels). Slit a-b shows the location through the prominence-corona system that is used to construct the space-time map of propagating intensity disturbances plotted in Fig.\,\ref{fig:pd2}. Slit c-d is a cut through prominence spicules that is used to plot wave motions in them in Fig.\,\ref{fig:wave2}. 
\textit{Panel d}: Root mean square intensity fluctuations at two different time periods (as noted in the panels) showing prominence spicules emanating from the prominence (see Appendix\,\ref{sec:rms} for details). The dotted line marks the location from where the RMS intensity is plotted as a function of distance along the slit in Fig.\,\ref{fig:wid}. An animation of \textit{panel b} is available online. See Sect.\,\ref{sec:space1} and Appendices\,\ref{sec:prom1} and \ref{sec:met} for details. 
\label{fig:cont}}
\end{center}
\end{figure*}

Skylab observations of solar prominences in the 1970s led to detailed comparisons of the prominence-corona interface with the chromosphere-corona transition region. Despite systematic differences between the two systems (e.g., differences in densities), these observations pointed to a striking similarity between the cool prominence threads and chromospheric spicules \citep{1976SoPh...50..365O}. At the prominence-corona interface, spectroscopic observations from SOHO/SUMER exhibit signatures of turbulent motions that are comparable to those found at the chromosphere-corona transition region. This led to suggestions that the prominence-corona interface is also heated in a similar way by MHD waves \citep{2004SoPh..223...95C,2007A&A...469.1109P}. Although quiescent condensations exhibit waves at lower temperatures of $10^4$\,K \citep{2009ApJ...704..870L,2013ApJ...779L..16H,2013ApJ...777..108S}, evidence for heating signatures at higher coronal temperatures that would explain the prominence-corona transition region is rare. 

The heating imprints its signature on the optically thin extreme-ultraviolet (EUV) radiation emitted by plasma at coronal temperatures. It is nontrivial to extract such heating signatures from prominences at EUV because condensations emit only weakly at these wavelengths \citep{2012ApJ...754...66P}. For the most part, they appear dark in the EUV due to bound-free absorption in hydrogen and helium Lyman continua \citep{1998SoPh..183..107K}. Observations could be impeded by the line-of-sight EUV emission. Therefore, even the serendipitous cases with little confusion from the back- and foreground emission will provide information on the processes responsible for the heating of the prominence-corona interface, which plays an important role in the long-term stability of condensations. 

\begin{figure*}
\begin{center}
\includegraphics[width=120mm]{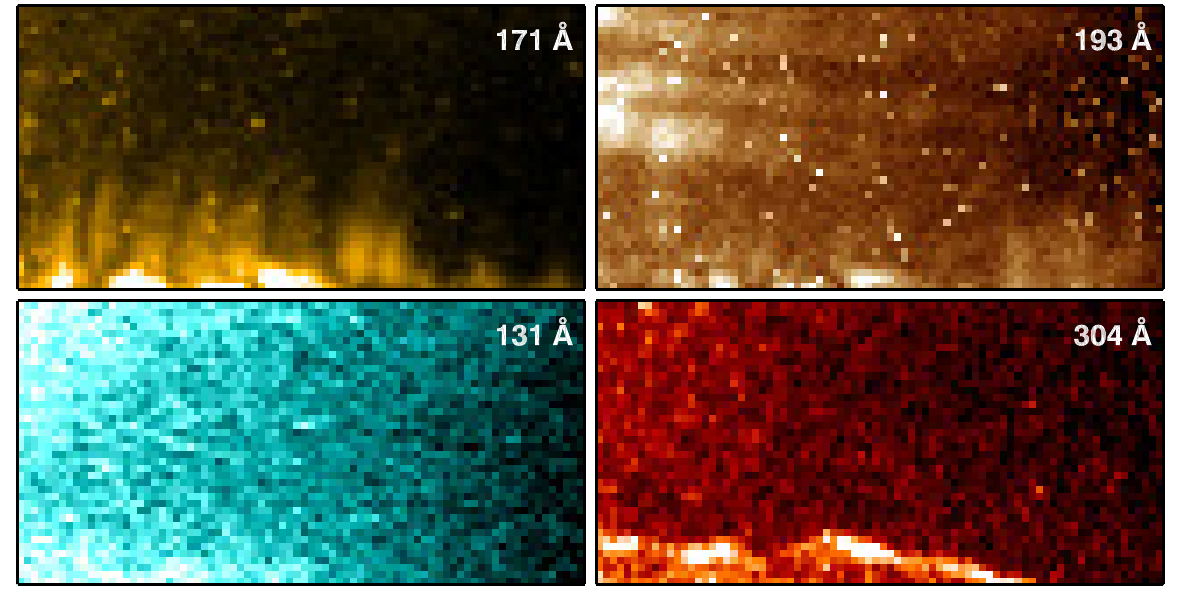}
\caption{Thermal structuring of prominence spicules. The top left panel is same as the top panel of Fig.\,\ref{fig:cont}d. The other panels display the maps of RMS intensity fluctuations obtained from the AIA 193\,\AA, AIA 131\,\AA, and AIA 304\,\AA\ filter images during 2015 February 15 UT 20:00-20:29. See Sect.\,\ref{sec:space1}.
\label{fig:temp}}
\end{center}
\end{figure*}

Here we investigate the dynamics at the prominence-corona transition region of two quiescent prominences that were observed with the Atmospheric Imaging Assembly \citep[AIA;][]{2012SoPh..275...17L,2012SoPh..275...41B} on board the Solar Dynamics Observatory \citep[SDO;][]{2012SoPh..275....3P}. These structures are less strongly affected by the back- and foreground EUV emission. A closer examination of the interface region revealed heated jets that we refer to as ``prominence spicules'', which emanate from the turbulent motions of the prominence material. These prominence spicules imprint their signature on the emission as quasi-periodic intensity disturbances that propagate into the corona and show signatures of transversal Alfv\'{e}nic waves. Our results point to the crucial role of turbulence-driven heating in the evolution and stability of solar condensations. Such a magnetized turbulence is also generic to filaments in the interstellar medium \citep{2013A&A...556A.153H}.

\section{Prominence-corona transition region}

Here we focus on a prominence observed on 2015 February 14 and 15 (Figs.\,\ref{fig:cont}a and c; see Appendix\,\ref{sec:prom1} for observational details; in Appendix\,\ref{sec:prom2} we discuss another case of a prominence observed in 2016). Based on a visual inspection of the prominence evolution, we find that this interface exhibits pervasive motions on both days. Moreover, the surface of the interface region is clearly distinguishable. The AIA 171\,\AA\ EUV filter, with contributions from the \ion{Fe}{ix} line (equilibrium temperature of 0.7\,MK) and \ion{Fe}{x} (equilibrium temperature of 1.1\,MK), detects faint emission from prominences \citep{2012ApJ...754...66P}. This is also the case with our example. In the following, we present the properties and dynamics of the prominence-corona transition region.

\begin{figure}
\includegraphics[width=88mm]{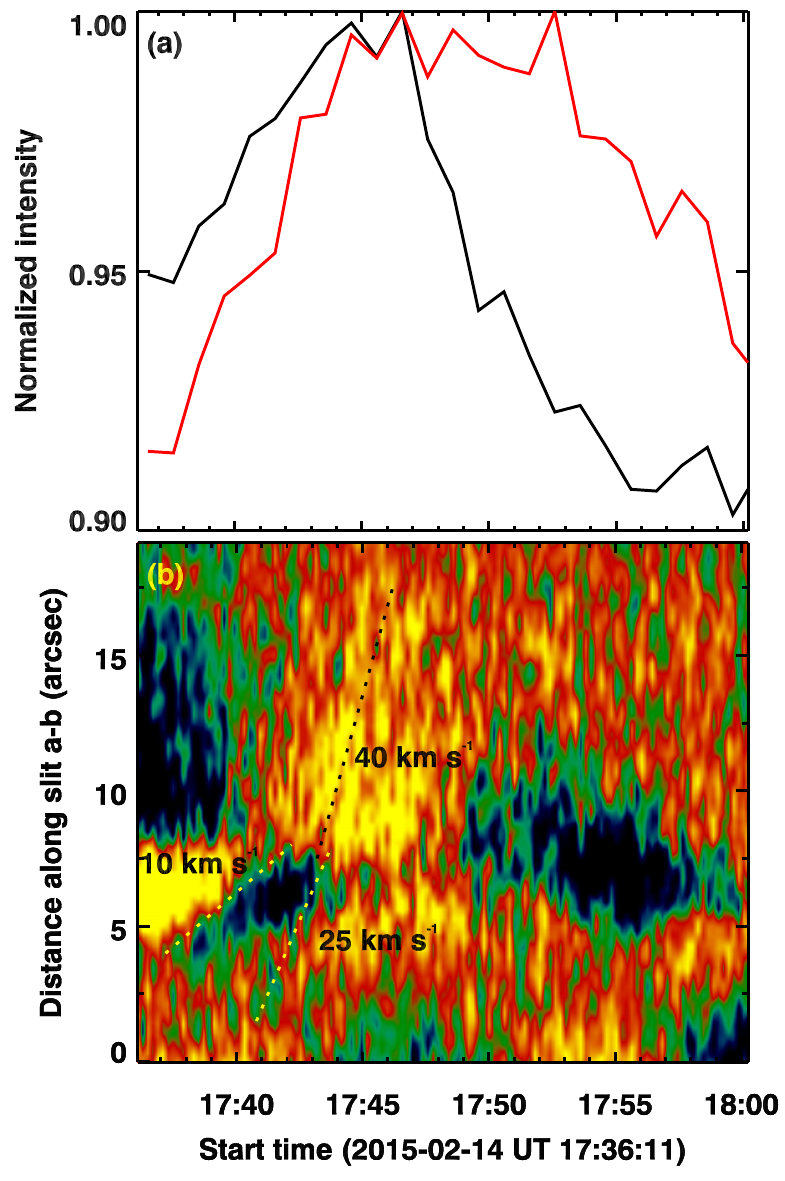}
\caption{Generation and propagation of a prominence spicule. \textit{Panel a}: Intensity vs. time at two locations along the spicule marked in Fig.\,\ref{fig:cont}b. The two light curves are average intensities obtained from the same colored regions as are shown in Fig.\,\ref{fig:cont}b. \textit{Panel b}: Space-time map from slit a-b along the spicule shown in Fig.\,\ref{fig:cont}b. The slanted dotted lines are for reference to indicate the propagating speeds. See Sect.\,\ref{sec:dyn} and Appendix\,\ref{sec:pdw} for details.
\label{fig:pd1}}
\end{figure}

\subsection{Spatial and thermal structure} \label{sec:space1}

Upon a closer look, and in particular, following the evolution of the prominence-corona interface on 2015 February 14, the faint emission is seen to be richly structured with distinguishable needle-like jet features or prominence spicules (Fig.\,\ref{fig:cont}b; Movie 1). From these maps we identified that the prominence spicules are anchored at the apparent surface of the prominence, and in this case, extend $\sim$15\arcsec\ or more into the surrounding corona, beyond which they fade away in the background emission. The faint EUV emission from these prominence spicules can also be detected in the root mean squared (RMS) maps (see Appendix\,\ref{sec:rms} for details). Similar to the prominence spicules seen in direct imaging (Fig.\,\ref{fig:cont}b), these RMS maps also show distinguishable spicular structures at the prominence-corona interface (Fig.\,\ref{fig:cont}d).

The spatial structuring of these prominence spicules shows that they have a width of more than 2\arcsec (see Appendix\,\ref{sec:space2}). Their apparent orientation is almost locally normal to the prominence surface, following the helical magnetic field, and they lie roughly parallel to the solar surface. Similar spicules are seen throughout the prominence-corona interface on the northern side. 

To investigate the thermal structuring of the prominence spicules, we created maps of RMS intensity fluctuations from the other filters, that is, AIA 193\,\AA, AIA 131\,\AA, and AIA 304\,\AA\ (see Appendix\,\ref{sec:rms}). These maps along with the top panel of Fig.\,\ref{fig:cont}d are displayed in Fig.\,\ref{fig:temp}. None of the RMS maps except for the AIA 171\,\AA\ case display any clear spicular structuring. Because they are absent in the 131\,\AA\ and 304\,\AA\ maps (both filters are sensitive to emission from cooler plasma, Appendix\,\ref{sec:obs}), it is likely that the prominence spicules lack a cooler plasma component below 0.7\,MK. However, these observations might be limited by the temperature sensitivity of the AIA.

\begin{figure*}
\begin{center}
\includegraphics[width=180mm]{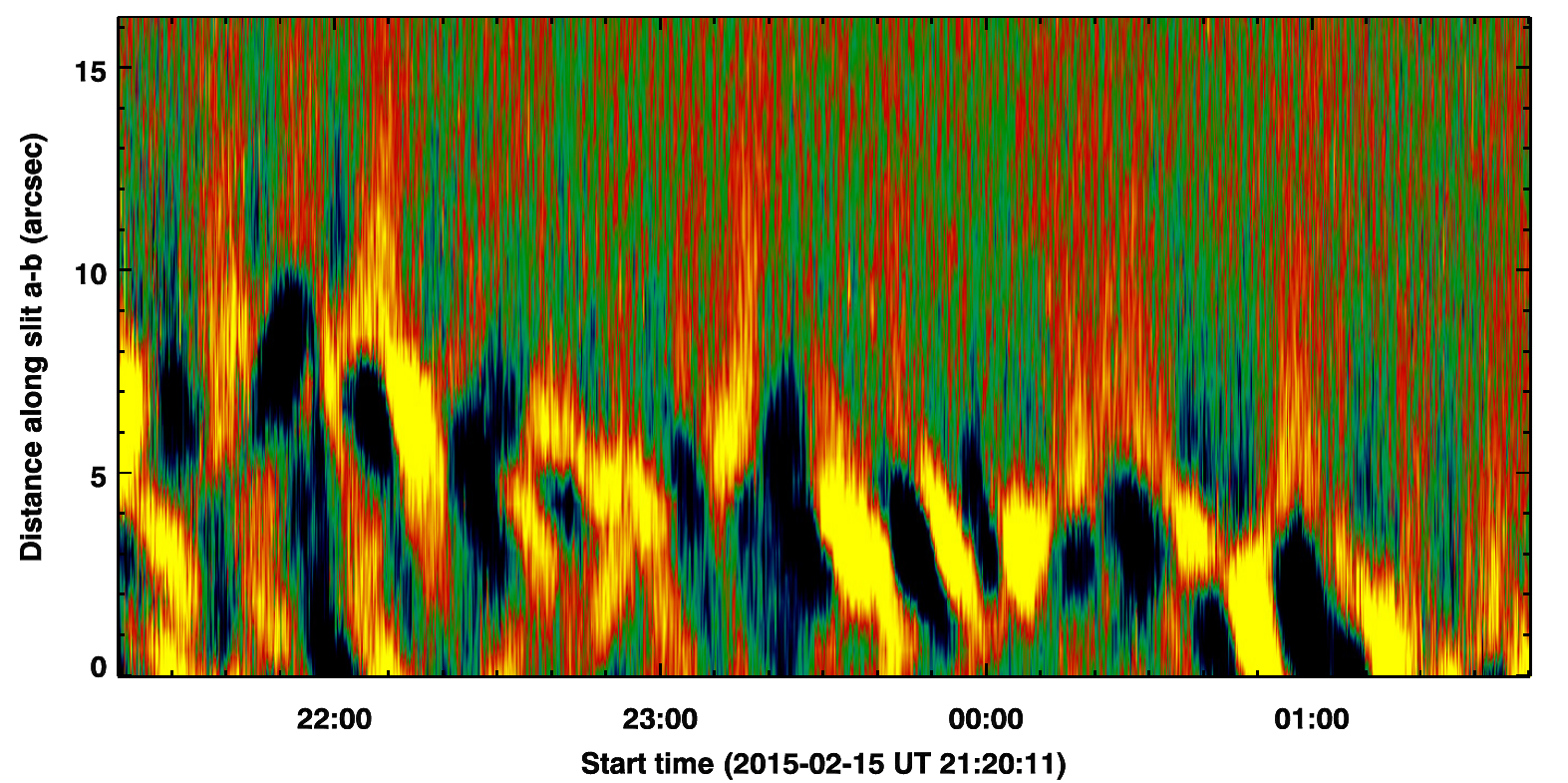}
\caption{Long-term evolution of the prominence spicules exhibiting quasi-periodic propagating intensity disturbances. Similar to Fig.\,\ref{fig:pd1}, but plotted for the observations obtained on 2015 February 15 along slit a-b in Fig.\,\ref{fig:cont}c. See Sect.\,\ref{sec:dyn} and Appendix\,\ref{sec:pdw} for details.
\label{fig:pd2}}
\end{center}
\end{figure*}

\subsection{Dynamics} \label{sec:dyn}

At the footpoints of the prominence spicules, the prominence surface exhibits spatial corrugations that change with time. These changes are seen as rapid out- and inward surge-like motion of the prominence material itself, with prominence spicules emanating from the tip of the corrugations (for reference, the dotted white line is overlaid in Fig.\,\ref{fig:cont}b to visualize these turbulent motions; see also Movie 1). We illustrate these dynamics with an example (marked with a slit as a dotted black line in Fig.\,\ref{fig:cont}b). We display the intensity along the slit as a function of time in Fig.\,\ref{fig:pd1}b. The prominence material surges with speeds in the range of 10\,km\,s$^{-1}$ to 25\,km\,s$^{-1}$. The prominence spicule then extends from the tip of this prominence surge, and the intensity disturbance associated with it propagates outward into the corona with speeds of $\sim$40\,km\,s$^{-1}$. The whole event lasted for about 10\,minutes. Light curves from two locations along that prominence spicule demonstrate the propagating nature of the intensity disturbance because the emission from a region on the spicule closer to the prominence leads the emission farther away along the same structure (Fig.\,\ref{fig:pd1}a). These prominence surges and spicules, each lasting for only a few minutes, are repeated quasi-periodically over hours (Fig.\,\ref{fig:pd2}, see also Appendix\,\ref{sec:prop}). Because their propagation is roughly parallel to the solar surface, gravity does not play a role in their dynamics. This distinguishes them from the dynamics of chromospheric spicules that mostly propagate radially into the corona, under the influence of the Sun's gravity.

\begin{figure*}
\begin{center}
\includegraphics[width=120mm]{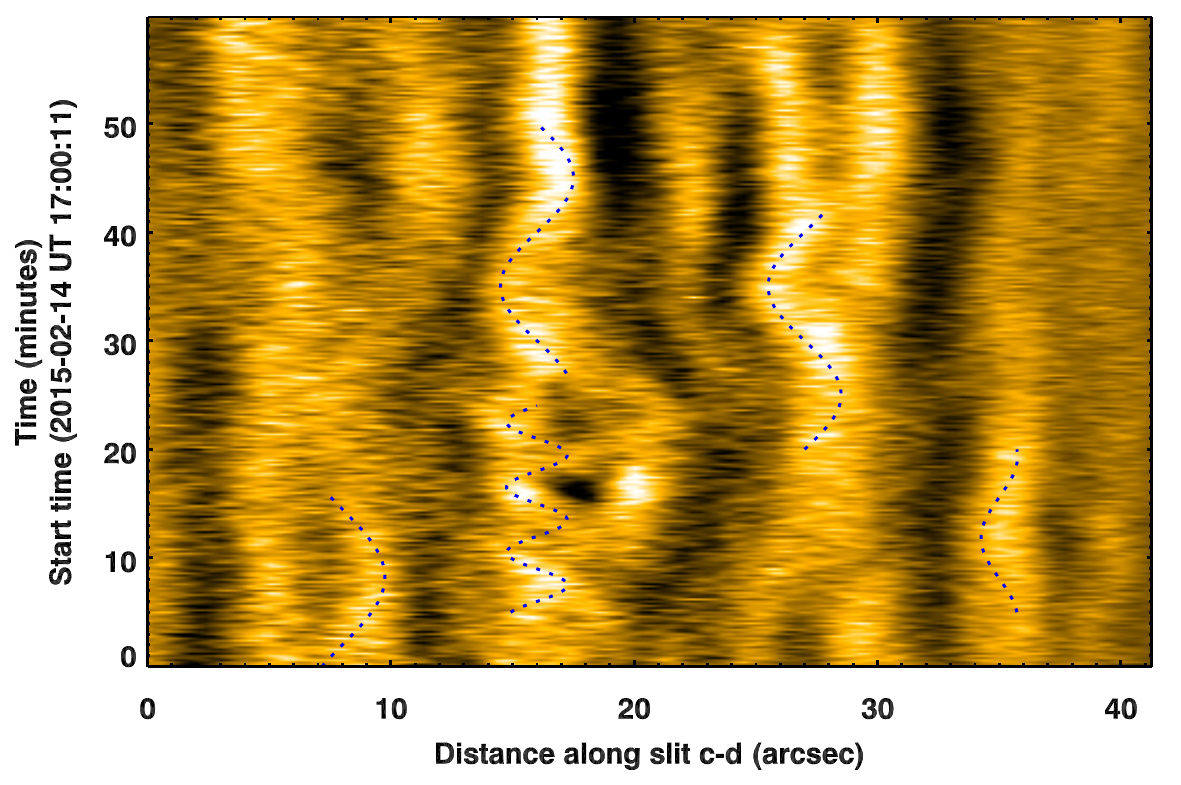}
\caption{Transverse oscillations exhibited by prominence spicules. The smooth-subtracted image shows the sporadic transverse Alfvénic waves exhibited by prominence spicules, crossing slit c-d in Fig.\,\ref{fig:cont}b. The dotted curves are sine waves overlaid on spicule oscillations to guide the eye. See Sect.\,\ref{sec:dyn} and Appendix\,\ref{sec:pdw} for details.
\label{fig:wave1}}
\end{center}
\end{figure*}

In addition to channeling propagating intensity disturbances, the prominence spicules display sporadic transverse oscillations. These oscillatory motions of prominence spicules are shown in Fig.\,\ref{fig:wave1}, with sine waves overlaid to guide the eye (see also Appendix\,\ref{sec:prop}). These transverse waves exhibited by the prominence spicules offer diagnostics on the extended magnetic fields of prominences by means of seismological techniques \citep{2018LRSP...15....3A}. In this case, the apparent wave amplitudes in prominence spicules are in the range of 0.5 to 2.5\,km\,s$^{-1}$. This is an order of magnitude smaller than the amplitudes sustained by chromospheric spicules \citep{2011Natur.475..477M}; consequently, the energy flux carried by these transverse waves is insufficient to heat the prominence-corona transition region.

\section{Discussion and conclusions}
The energy that is required to heat and launch the prominence spicules (and to excite MHD waves) could be supplied by turbulent motions. Using the event in Fig.\,\ref{fig:pd1} as a prototype, we provided an order-of-magnitude estimate for the energy flux. By this we determined whether these turbulent motions are able to provide enough energy to heat and propel the observed prominence spicules. The kinetic energy flux, $E$, associated with a moving parcel of fluid, is given by $E=0.5\rho u^3$. Here $\rho$ is the fluid density and $u$ is the fluid velocity. Using the canonical values of the prominence density (10$^{-11}$ to 10$^{-10}$\,kg\,m$^{-3}$) \citep{2010SSRv..151..243L} and the observed speeds of the prominence surge (10 to 25\,km\,s$^{-1}$), we obtained an energy flux of 10\,W\,m$^{-2}$ to 1000\,W\,m$^{-2}$. This is in the range of energy fluxes that are sufficient to generate and sustain a quiescent corona \citep{1977ARA&A..15..363W,2014ApJ...790...68K} around a prominence.

Prominence surges, spicules, quasi-periodic propagating intensity disturbances, and sporadic waves through the prominence-corona interface are all similar to those that are widely spread in the regular chromosphere-corona system, which is ultimately powered by granular motions on the solar surface. Numerical models suggest that chromospheric spicules and related dynamics through the chromosphere corona occur when the magnetic tension is impulsively released through the interaction between neutrals and ions (or ambipolar diffusion) \citep{2017Sci...356.1269M}. Ion-neutral interactions are also prevalent in the interface region of a prominence-corona system and they give rise to turbulent motions \citep{2014A&A...565A..45K}. Here we provided observational evidence that these turbulent motions lead to heated prominence spicules at the prominence-corona transition region. 

Recent studies identified dynamics such as the transient ejection of plasma blobs \citep{2013ApJ...770...35S} or rotational motions \citep{2012ApJ...752L..22L} along the helical magnetic fields of solar prominences, but their source remains unclear. The prominence spicules that channel propagating disturbances and jets and the underlying turbulent motions are potential candidates and sources for explaining these dynamics. According to our conservative estimates, the prominence spicules evaporate a prominence such as we discussed in Fig.\,\ref{fig:cont} in a matter of some days to a few weeks (see Appendix\,\ref{sec:evap}). These timescales are short compared to the lifetime of that filament (cf. Appendix\,\ref{sec:prom1}). This provides a new possible process for the decay of filaments. These prominence spicules are not predicted by the current MHD models of prominences \citep{2016ApJ...823...22X}. Consequently, these models do not include the role of prominence spicules in the filament evolution, but the spicules might have a significant impact on the filament evolution. Overall, our observations revealed heated prominence spicules emanating from cooler filaments, which are solar analogs to the magnetized and turbulent plasma condensations that are widely observed in astrophysics.

\begin{acknowledgements}
We gratefully acknowledge the constructive and detailed comments by the referee,\,Eric Priest, that helped us to improve the paper. L.P.C. acknowledges previous funding from the European Union's Horizon 2020 research and innovation program under the Marie Sk\l{}odowska-Curie grant agreement No.\,707837. L.L. is supported by the National Natural Science Foundations of China (11673034 and 11533008). SDO data are courtesy of NASA/SDO and the AIA and HMI science teams. This research has made use of NASA's Astrophysics Data System. We acknowledge the usage of JHelioviewer software \cite[][]{2017A&A...606A..10M}.
\end{acknowledgements}


\begin{appendix}

\section{Observations} \label{sec:obs}

To study the dynamics of the plasma condensations in the solar corona, we considered high-cadence observations of two quiescent prominences observed in the EUV with SDO/AIA. In particular, we concentrated on observations from the AIA 171\,\AA, 193\,\AA, 131\,\AA,\ and 304\,\AA\ filters. We used level-1 AIA data in this study. The EUV images have a plate scale of 0.6\arcsec pixel$^{-1}$ and cadence of 12\,s \citep{2012SoPh..275...41B}. The data were then tracked to remove the solar rotation and were coregistered to a common field of view using publicly available standard IDL procedures in the solarsoft maps package. 

Here we briefly summarize the thermal characteristics of the AIA filters \citep{2012SoPh..275...41B}. The AIA 171\,\AA\ filter has a dominant contribution to emission from \ion{the Fe}{ix} line that forms at an equilibrium temperature of 0.7\,MK and also detects emission from\ion{ the Fe}{x} line, which forms at 1.1\,MK. The AIA 193\,\AA\ filter receives a dominant contribution to emission from the \ion{Fe}{xii} line at $\sim$1.5\,MK, but it also receives contribution from the \ion{O}{v} line that forms at $\sim$0.3\,MK. Under non-flaring conditions, the AIA 131\,\AA\ filter receives contribution from the \ion{Fe}{viii} line that forms at $\sim$0.4\,MK and from the \ion{O}{vi} line that forms at $\sim$0.3\,MK. The AIA 304\,\AA\ filter is dominated by contribution from \ion{He}{ii} line, which forms in the temperature range of 0.05\,MK to 0.1\,MK.

\subsection{Prominence of 2015} \label{sec:prom1}

Snapshots of the selected 2015 prominence as seen with the AIA 171\,\AA\ filter are displayed in Figs.\,\ref{fig:cont}a and c. This is a long-lived structure. This condensation first appeared on 2015 January 9 near the southeast limb. It remained distinguishable in the AIA 171\,\AA\ filter images for at least five solar rotations ($\sim$20\,weeks), until 2015 June 1. We used observations on two consecutive days when a large portion of this prominence was observed to be suspended over the southwest limb on 2015 February 14 and 15. During these days. the prominence was oriented along its axis at the limb, providing the clearest view of its interface with the corona. There are no apparent EUV structures (e.g., coronal loops) in the line of sight, in particular at the northern prominence-corona interface, which is our region of interest for further analyses.

\subsection{Prominence of 2016} \label{sec:prom2}

\begin{figure*}
\begin{center}
\includegraphics[width=180mm]{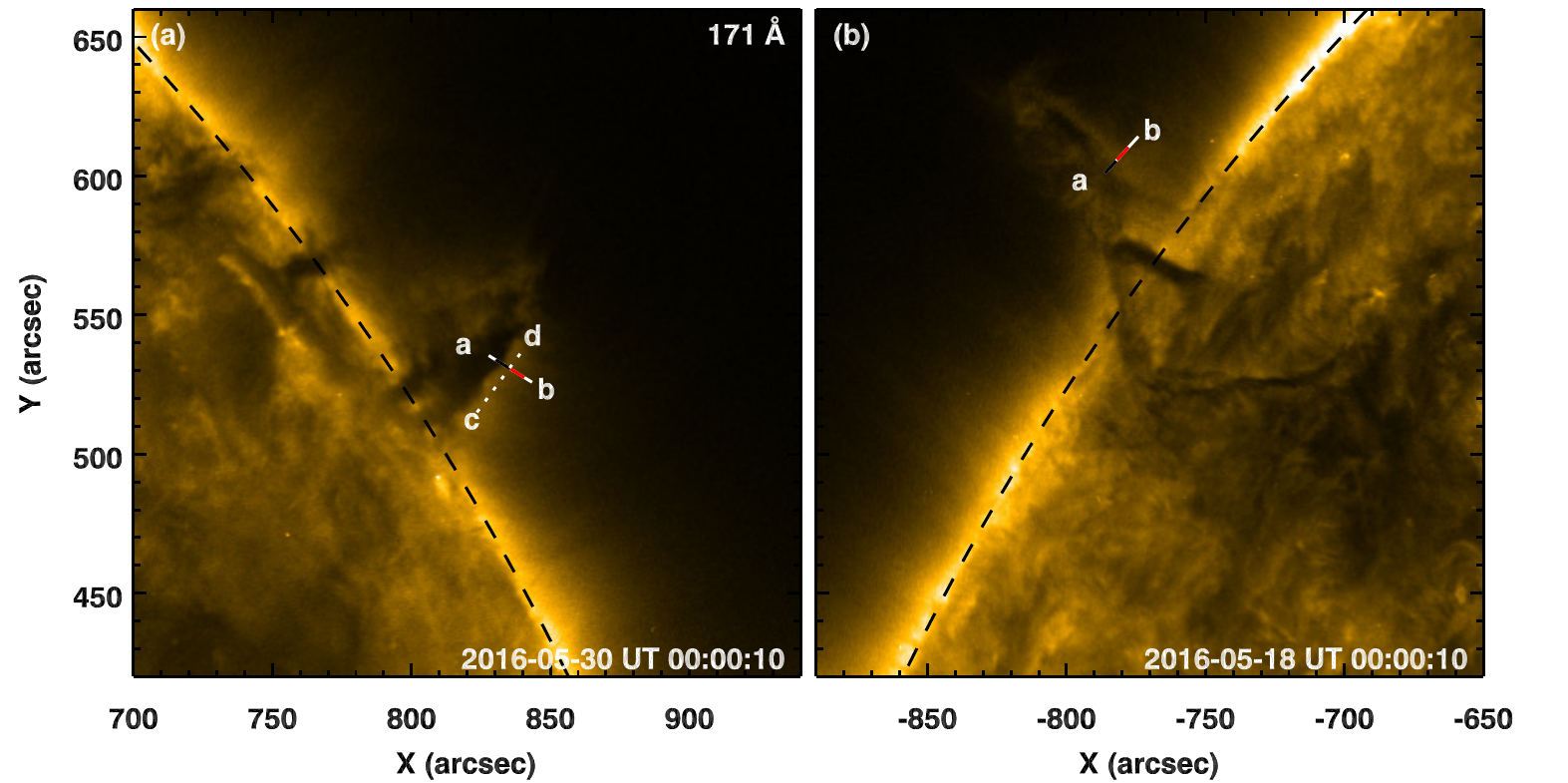}
\caption{Imaging observations of a prominence-corona system 12 days apart. Same as Fig.\,\ref{fig:cont}a, but plotted for the observations obtained on 2016 May 30 (panel a) and 2016 May 18 (panel b). Slits a-b (in both panels) are used to construct the respective space-time displayed in the lower panels of Figs.\,\ref{fig:pd3} and \ref{fig:pd4}. The black and red marked regions along each slit are used to construct the average AIA 171\,\AA\ emission that is plotted as the same color-coded light curves as in the top panels of Figs.\,\ref{fig:pd3} and \ref{fig:pd4}. Slit c-d in \textit{panel a} is used to plot wave motions in prominence spicules in Fig.\,\ref{fig:wave3}. See Appendix\,\ref{sec:prom2}. 
\label{fig:cont2}}
\end{center}
\end{figure*}

The 2015 prominence has a clearly distinguishable and continually evolving interface with the surrounding corona. To further investigate the signatures of prominence spicules, we selected another example of a prominence observed in 2016, which has a more highly diffused interface with the corona that is not clearly distinguishable (Fig.\,\ref{fig:cont2}), in contrast to the 2015 case. Based on a visual inspection, we find that the evolution of the prominence-corona interface of this second example is not as pervasive as the 2015 example. This second prominence first appeared near the northeast limb on 2016 March 23 in the AIA 171\,\AA\ filter. It remained distinguishable for at least three solar rotations ($\sim$13\,weeks), until 2016 June 27. Therefore, similar to the first case, this condensation is also a long-lived structure. We used AIA observations that are 12\,days apart (roughly half a solar rotation), when the prominence appeared off-limb in the eastern and western hemispheres. In particular, we focused on regions over the northeast limb (2016 May 18) and northwest limb (2016 May 30). There are no apparent EUV structures in the line of sight. The faint EUV emission surrounding this prominence is less strongly structured and diffused (i.e., smoother) than the first case. Despite this difference, propagating disturbances and waves are detected at its interface with the corona (Figs.\,\ref{fig:pd3}, \ref{fig:pd4}, and \ref{fig:wave3}).

\section{Methods} \label{sec:met}

To reveal the dynamics of prominence spicules emanating from prominences in the EUV images, we employed common image processing techniques as described below. 

\subsection{Detecting faint prominence spicules using RMS maps}\label{sec:rms}
In Fig.\,\ref{fig:cont}b we display a direct image of prominence spicules. In general, however, because of the faint nature of the EUV emission at the prominence-corona interface, it is difficulte to directly observe spicules over the background emission. To mitigate this issue, we made use of the fact that these prominence spicules exhibit propagating intensity disturbances along their length (cf. Sect.\,\ref{sec:dyn}). When observed as a function of time, these disturbances can be used to recover the underlying spicular feature. To this end, we derived RMS intensity maps from the original images. These maps will reveal small intensity changes and thus the faint prominence spicules we investigate. The RMS maps recover fluctuations with respect to the average intensity over a given period of time. Let $I_\text{avg} (x,y)$ be the intensity map obtained by averaging $n$ original images. Then the map of the RMS intensity, $I_\text{rms} (x,y)$, is defined as
\begin{equation}
I_\text{rms} (x,y) = \sqrt{\frac{\sum\limits_{i=1}^{n}[I(x,y,i)-I_\text{avg} (x,y)]^2}{n}},
\end{equation}
where $I(x,y,i)$ is the $i$th image in the time series. For our purpose, we selected $n=150$ consecutive images. The image cadence of 12\,s of AIA 171\,\AA\ images implies that a single RMS map carries information from 30\,minutes (for $n=150$). These maps, including the region of interest and the corresponding time period, are displayed in Figs.\,\ref{fig:cont}d and \ref{fig:temp}.

\subsection{Detecting propagating disturbances and waves using space-time maps}\label{sec:pdw}

The prominence spicules are associated with propagating intensity disturbances and transverse MHD waves (see Figs.\,\ref{fig:pd1}, \ref{fig:pd2}, and \ref{fig:wave1}). These dynamics are best seen in the space-time maps that capture intensity variation along a given direction or cut as a function of time. We used AIA 171\,\AA\ filter images to construct such maps.

To investigate any propagating disturbances along the prominence spicules, we used slits along selected directions (slits a-b in Figs.\,\ref{fig:cont} and \ref{fig:cont2}). We interpolate the intensity along these slits and averaged the intensity from $\pm$1.2\arcsec\ across the slit to improve the signal. The interpolated arrays of intensity along the slit were then stacked together as a function of time to create space-time maps. To enhance the signal contrast in the temporal direction, we first smoothed the space-time maps with a time window of 20\,minutes (corresponding to 100 snapshots). Next, these smoothed maps were subtracted from the original maps to create contrast-enhanced space-time maps. These contrast-enhanced maps are displayed in Figs.\,\ref{fig:pd1}b, \ref{fig:pd2}, \ref{fig:pd3}, and \ref{fig:pd4}. The propagating disturbances along the slits in the plane of sky can be seen as quasi-periodic (slanted) intensity enhancements in these images. In Fig.\,\ref{fig:pd1}b reference lines are overlaid to guide the eye.

The light curves in Figs.\,\ref{fig:pd3} and \ref{fig:pd4} were obtained by averaging emission from arrays that were computed for the space-time maps. The spatial extent for averaging is indicated by the same colored bars that are overlaid along slits a-b in Fig.\,\ref{fig:cont2}. The light curves were then normalized to the maximum intensity during the plotted period. 


Filament threads exhibit transverse oscillations that are generally interpreted as MHD waves that are carried by the underlying magnetic field  \citep{2009ApJ...704..870L,2013ApJ...779L..16H,2015ApJ...809...71O}. To identify the presence of such transverse waves in the prominence spicules, we considered slits that cut through them (slits c-d in Figs.\,\ref{fig:cont} and \ref{fig:cont2}). We interpolated the intensity along these slits, and to improve the signal, we averaged the  intensity from $\pm$0.6\arcsec\ (for the 2015 case; Appendix\,\ref{sec:prom1}) or $\pm$3\arcsec\ (for the 2016 case; Appendix\,\ref{sec:prom2}) across the selected slit. These arrays of intensity along the slits were then stacked together as a function of time to create space-time maps. These maps were contrast-enhanced in the spatial direction with a window of $\sim$7\arcsec. The contrast-enhanced maps are displayed in Figs.\,\ref{fig:wave1}, \ref{fig:wave2}, and \ref{fig:wave3}. The brighter ridges seen in these maps correspond to the emission from prominence spicules that exhibit sporadic transverse oscillatory motions in the plane of sky. 

\subsection{Spatial maps of prominence spicules} \label{sec:mov}

The faint EUV emission that extends into the surrounding corona from the prominence-corona interface is structured into thin spicule-like features, particularly in the AIA 171\,\AA\ filter images (Fig.\,\ref{fig:cont}). For display purpose, we integrated the original 12\,s cadence data to one-minute-cadence time series to produce the snapshots in Fig.\,\ref{fig:cont}b and the associated online animation. The movie shows one-hour-long time series from the AIA 171\,\AA\ filter observations of the prominence observed on 2015 February 14, at a cadence of one\,minute. In the right panel of the movie we also show smooth-subtracted snapshots (that enhance the contrast of prominence spicules) from the left panel. To produce these enhanced images, we used a box of 3\arcsec$\times$3\arcsec\ to first spatially smoothen each snapshot in the time series. Next, these smoothed snapshots were subtracted from the unsmoothed time series, which resulted in a contrast-enhanced image sequence that better displays faint spicular structures. 

The movie highlights turbulent surge-like motions of the prominence material and the resulting prominence spicules. The slanted dotted line (same as in Fig.\,\ref{fig:cont}b) is plotted as a reference to visualize the out- and inward surge-like turbulent motions of the prominence material. To guide the eye, we also overlaid arrows that mark prominence surges (upward-pointing arrows) and spicules (downward-pointing arrows).

\section{Spatial structuring of prominence spicules} \label{sec:space2}

\begin{figure*}
\begin{center}
\includegraphics[width=180mm]{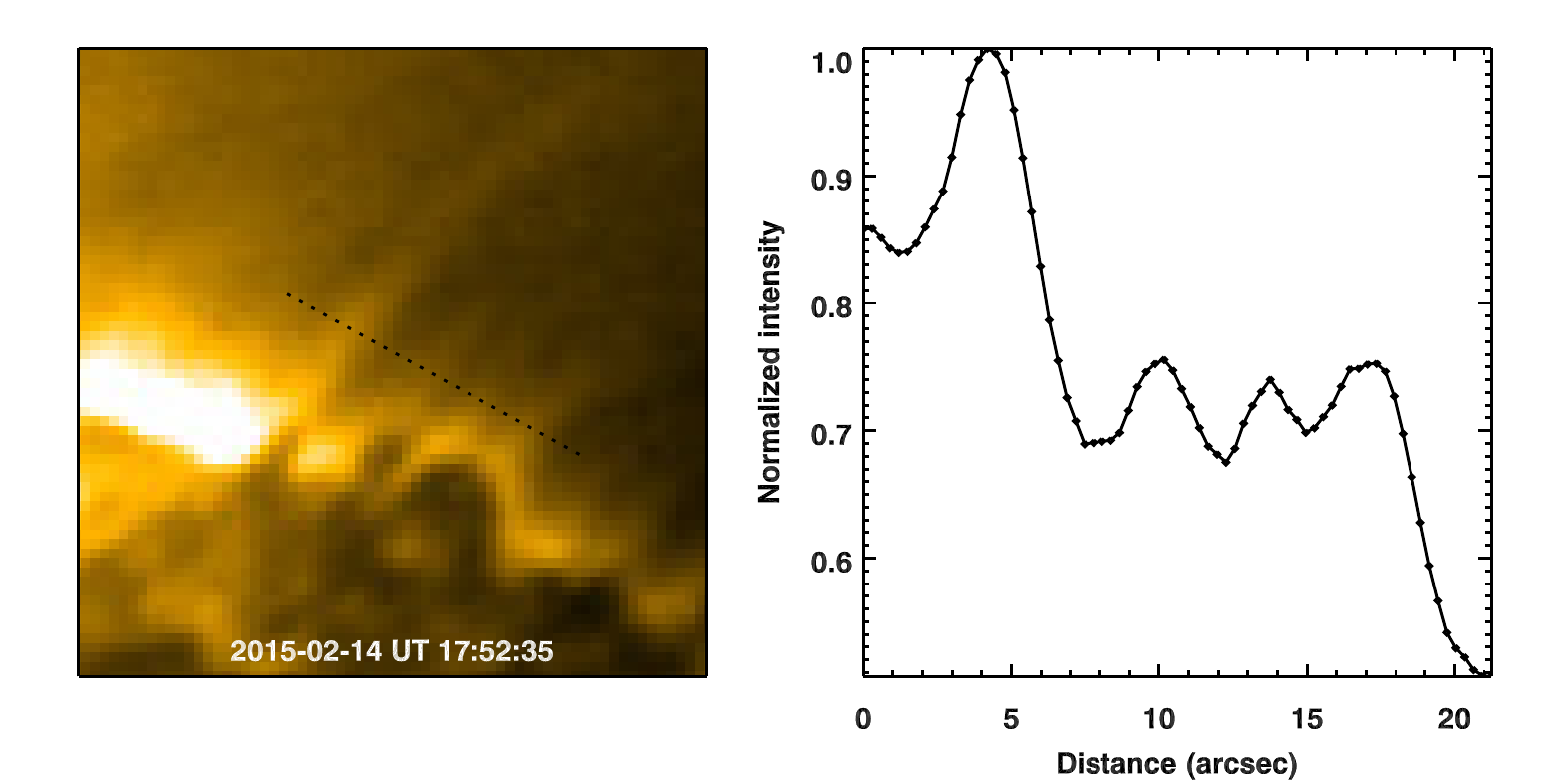}
\includegraphics[width=88mm]{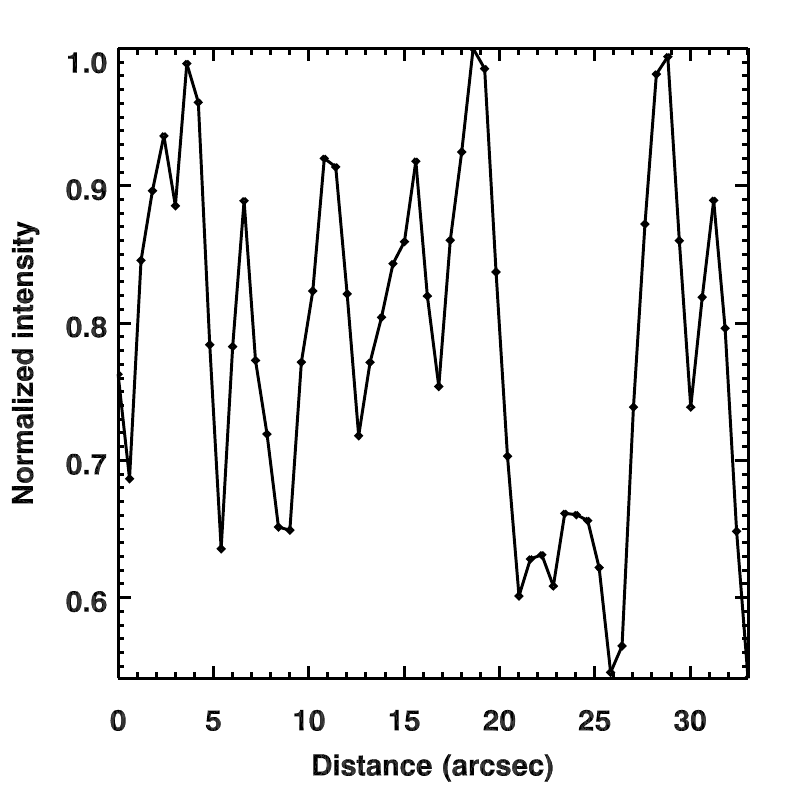}
\caption{Spatial structuring of prominence spicules. The top left panel is same as the bottom left panel in Fig.\,\ref{fig:cont}b. The black dotted line crosses a few structures at that instance. In the top right panel, we plot the AIA 171\,\AA\ intensity along the dotted line from the top left panel. The bottom panel shows the AIA 171\,\AA\ RMS intensity along the dotted line that is displayed in the top panel of Fig.\,\ref{fig:cont}d. See Appendix\,\ref{sec:space2} for discussion.
\label{fig:wid}}
\end{center}
\end{figure*}

In the case of the 2015 prominence (Sect.\,\ref{sec:obs}), prominence spicules are clearly visible in imaging observations (Fig.\,\ref{fig:cont}b) or through the proxy of RMS intensity fluctuations (Fig.\,\ref{fig:cont}d). These structures are distinguishable and have a finite width. To investigate the spatial structuring of these prominence spicules, we plot the AIA 171\,\AA\ intensities along selected slits that cross prominence spicules (dotted line in the top left panel of Fig.\,\ref{fig:wid}; dotted line in the top panel of Fig.\,\ref{fig:cont}d). The spatial structuring shows ridges with an intensity distribution (with widths of more than 2\arcsec) with clear peaks corresponding to the underlying prominence spicules (top right and bottom panels of Fig.\,\ref{fig:wid}). In this case, they are resolved in the AIA observations. These spicules at coronal temperatures are wider than the cooler ($\sim$10$^4$\,K) filament threads with widths of 0.3\arcsec\ to 0.4\arcsec\ \citep{2005SoPh..226..239L,2012A&A...541A.108V}. The widths of prominence spicules detected in the AIA 171\,\AA\ EUV images are comparable to the cross-sectional widths of coronal loops \citep{2017ApJ...840....4A}. It is possible that there are spicular structures below the resolution limit of AIA that give rise to a much smoother prominence-corona transition region (e.g., as in the case of the 2016 prominence; Fig.\,\ref{fig:cont2}). Future high-resolution EUV imaging or spectroscopic observations are required to comment on the unresolved prominence-corona transition region.

\section{Propagating disturbances and waves} \label{sec:prop}

\begin{figure*}
\begin{center}
\includegraphics[width=120mm]{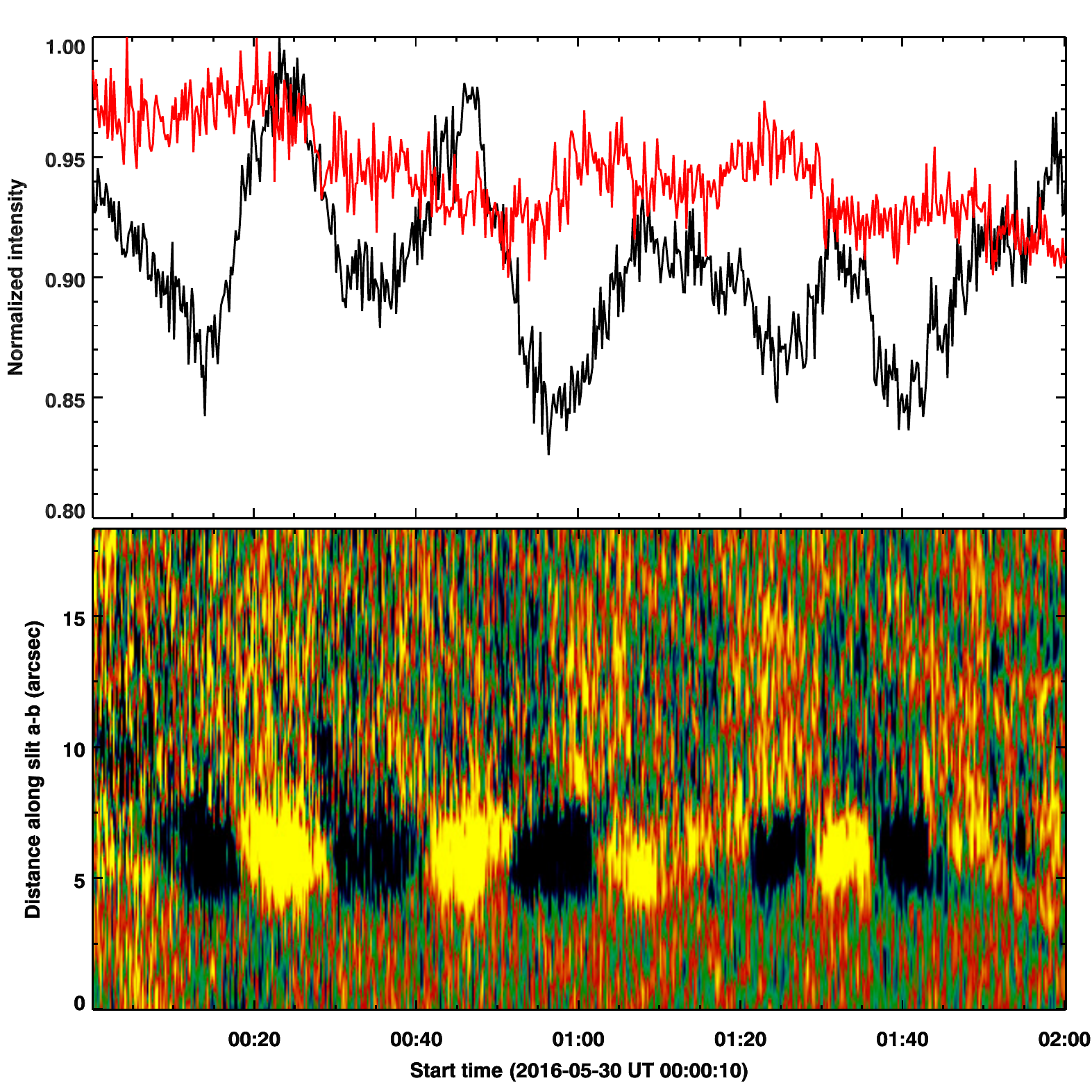}
\caption{Long-term evolution of the prominence spicules that exhibit quasi-periodic propagating intensity disturbances in a diffused prominence. Plotted for the observations obtained on 2016 May 30 (see Fig.\,\ref{fig:cont2}a). The top panel shows AIA 171\,\AA\ light curves averaged along the sections of the same colored slits as in Fig.\,\ref{fig:cont2}a. The bottom panel is the space-time map displaying propagating disturbances along slit a-b. See Appendices\,\ref{sec:pdw} and \ref{sec:prop}.
\label{fig:pd3}}
\end{center}
\end{figure*}

\begin{figure*}
\begin{center}
\includegraphics[width=120mm]{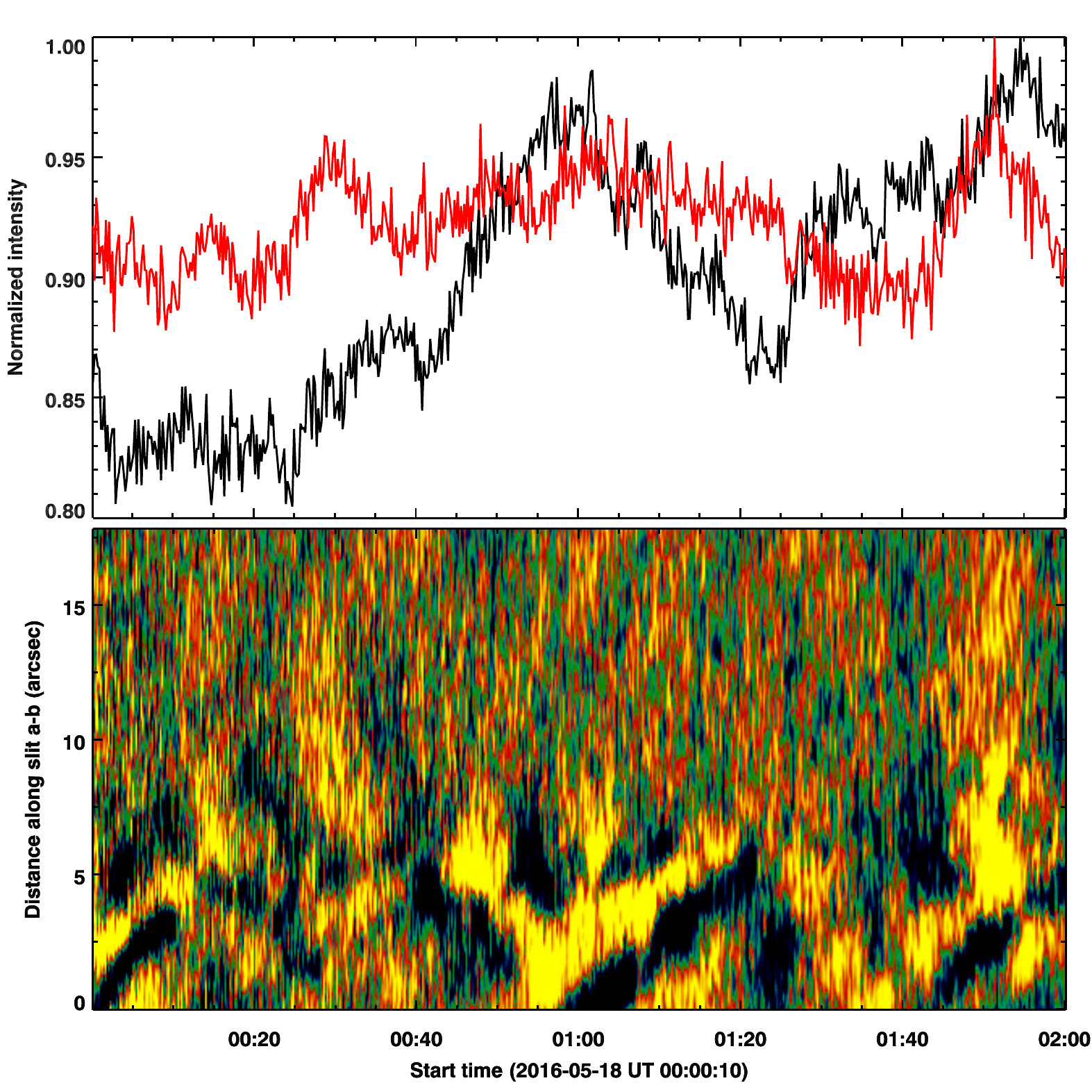}
\caption{Long-term evolution of the prominence spicules that exhibit quasi-periodic propagating intensity disturbances in a diffused prominence. Same as Fig.\,\ref{fig:pd3}, but plotted for the observations obtained on 2016 May 18. See Appendices\,\ref{sec:pdw} and \ref{sec:prop}.
\label{fig:pd4}}
\end{center}
\end{figure*}

In Sect.\,\ref{sec:dyn} we presented  in Fig. 3 the case of one selected propagating disturbance associated with a prominence spicule that lasted for about 10 minutes. We also find that disturbances like this are repeated quasi-periodically over several hours (Fig.\,\ref{fig:pd2}). Although its nature is more diffused than that in the 2015 case, the prominence observed in 2016 also exhibited propagating disturbances (Figs.\,\ref{fig:pd3}, and \ref{fig:pd4}). While the quasi-periodic signal is clear from 2016 May 30 observations (Fig.\,\ref{fig:pd3}), it is scrambled in 2016 May 18 observations (Fig.\,\ref{fig:pd4}). This could be due to line-of-sight superposition of signal from multiple prominence spicules at different phases in their evolution. 

The propagating disturbances along prominence spicules that we detected in our study are generated from the turbulent motions of plasma condensations. The direction of propagation is perpendicular to the local surface of the prominence into the surrounding corona (almost parallel to the solar surface). These findings further suggest that morphologically, the dynamics at the prominence-corona interface are very similar to those observed in the chromosphere-corona transition region. Overall, these quasi-periodic features are clearly analogous to the well-known propagating disturbances associated with chromospheric spicules. These disturbances from chromospheric spicules are generally thought to be the signatures of magnetoacoustic waves \citep{1998ApJ...501L.217D} or quasi-periodic plasma outflows \citep{2010ApJ...722.1013D}. 

\begin{figure*}
\begin{center}
\includegraphics[width=120mm]{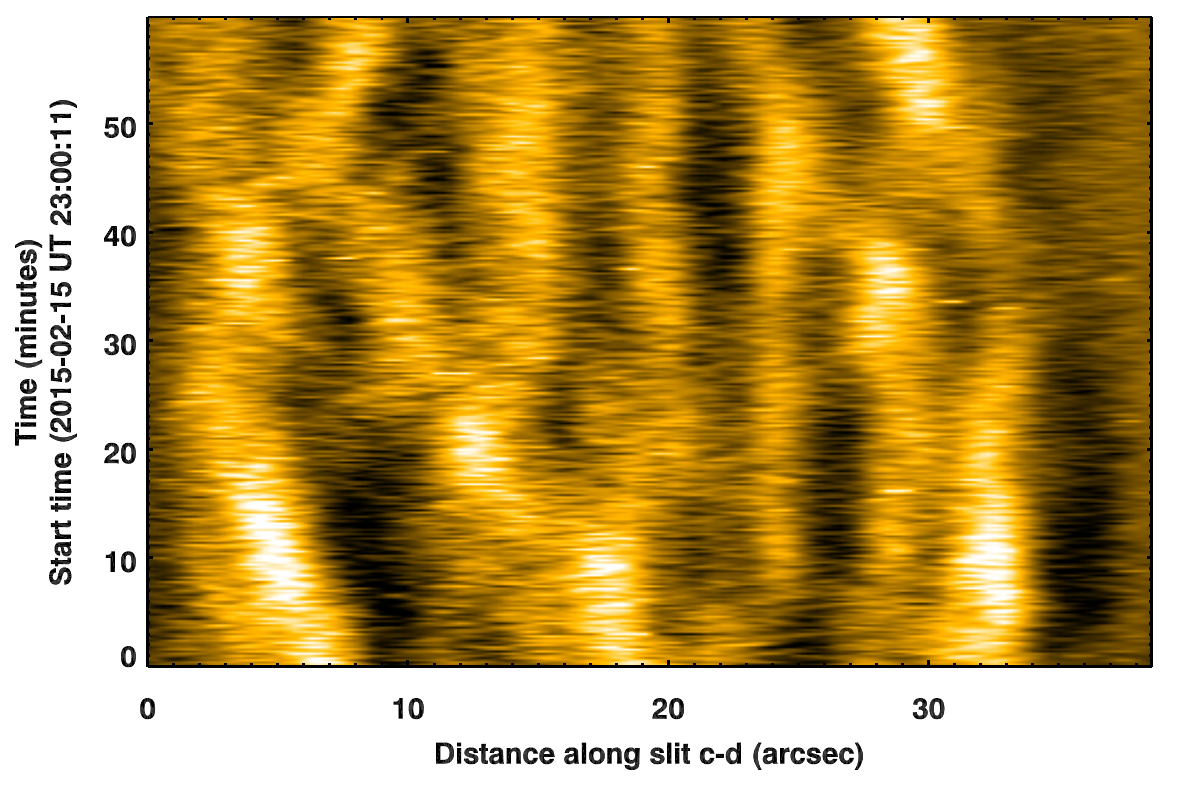}
\caption{Transverse oscillations exhibited by prominence spicules. The smooth-subtracted image shows the sporadic oscillations transverse to the spicules that cross slit c-d in Fig.\,\ref{fig:cont}c. See Appendices\,\ref{sec:pdw} and \ref{sec:prop}.
\label{fig:wave2}}
\end{center}
\end{figure*}

\begin{figure*}
\begin{center}
\includegraphics[width=120mm]{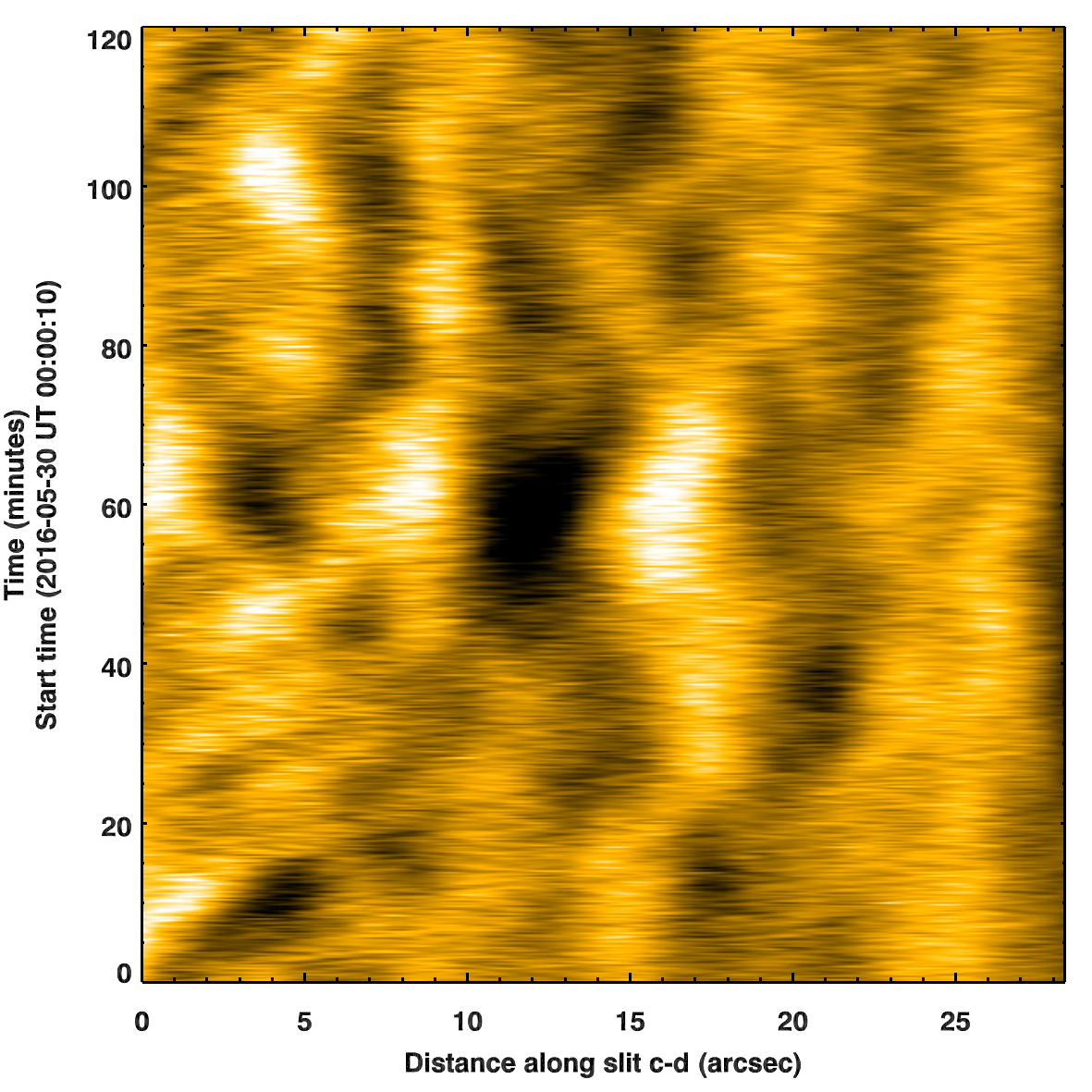}
\caption{Transverse oscillations exhibited by prominence spicules. The smooth-subtracted image shows the sporadic oscillations transverse to the spicules that cross the slit c-d in Fig.\,\ref{fig:cont2}a. See Appendices\,\ref{sec:pdw} and \ref{sec:prop}.
\label{fig:wave3}}
\end{center}
\end{figure*}

The wave motions exhibited by prominence spicules are not ubiquitous and are intermittent in both space and time (Figs.\,\ref{fig:wave1}, \ref{fig:wave2}, and \ref{fig:wave3}). It is of course possible that these prominence spicules have a component of wave motions out of the plane of the sky that is not captured by the imaging observations. High spatial resolution spectroscopic observations that enable velocity diagnostics through Doppler motions at coronal temperatures are required to retrieve these line-of-sight dynamics, but they are not available for our datasets.

\section{Filament evaporation due to prominence spicules} \label{sec:evap}

\begin{figure*}
\begin{center}
\includegraphics[width=120mm]{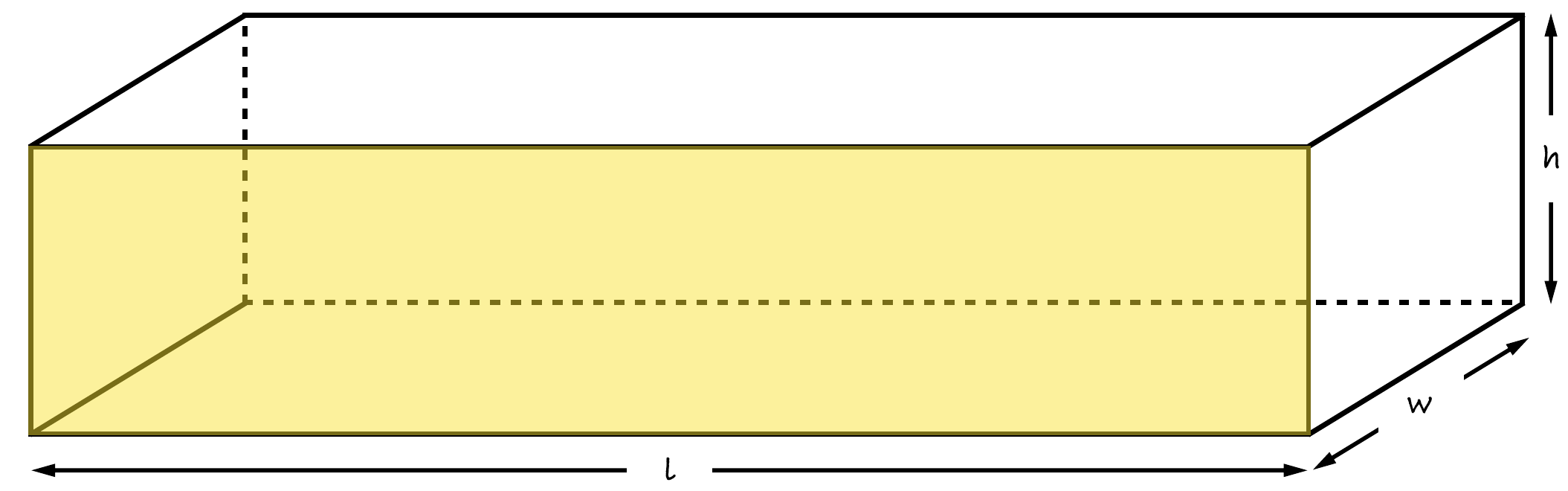}
\caption{Solar prominence represented with a cuboid. This prominence has a length ($l$), width ($w$), and height ($h$). The yellow shaded surface represents the region that is covered by prominence spicules (e.g., northern side of the prominence-corona system in Fig.\,\ref{fig:cont}). See Appendix\,\ref{sec:evap}.
\label{fig:box}}
\end{center}
\end{figure*}

The heated prominence spicules evaporate the prominence (or filament) material into the surrounding corona. The main question is whether this effect has the potential to affect the long-term evolution and lifetime of a prominence. Here we present simplistic but conservative estimates on the timescales associated with such an evaporation process, mediated by prominence spicules, using the case of the 2015 prominence. For this purpose, we assumed that the prominence is represented by a cuboid with length ($l$), width ($w$), and height ($h$) (see Fig.\,\ref{fig:box}). It is filled with cooler condensations with density $\rho_\text{pro}$. Then the total mass of the prominence is given by $m_\text{pro}=\rho_\text{pro}lwh$. As discussed in Sect.\,\ref{sec:dyn}, the turbulent motions generate and propel heated prominence spicules into the surrounding corona. Let $v$ be the velocity with which the prominence spicules are launched, and $\rho_\text{cor}$ be their density (equal to the coronal density). The mass loss due to these prominence spicules through the prominence surface is given by $\dot{m}=f\rho_\text{cor}vA$. Here $f$ is the filling factor or the fraction of the surface area, $A$, covered by the prominence spicules at a given time. In the case of the 2015 prominence, on both days (i.e., February 14 and 15, Fig.\,\ref{fig:cont}), the prominence spicules are visible at least on the northern side. It is likely that they extend over the entire length and height of the prominence (as illustrated by the shaded surface in Fig.\,\ref{fig:box}), over the area $A=lh$. We ignored the prominence spicules that emanated from the remaining surfaces. 

The timescale, $\tau$, for the evaporation is then given by 
\begin{eqnarray}
\tau &=& m_\text{pro}/\dot{m}_\text{pro}, \nonumber \\
       &=& \frac{1}{f}\cdot\frac{\rho_\text{pro}}{\rho_\text{cor}}\cdot\frac{lwh}{lh}\cdot\frac{1}{v}, \nonumber \\
       &=& \frac{1}{f}\cdot\frac{\rho_\text{pro}}{\rho_\text{cor}}\cdot w\cdot\frac{1}{v}.   
\end{eqnarray}

\begin{figure*}
\begin{center}
\includegraphics[width=120mm]{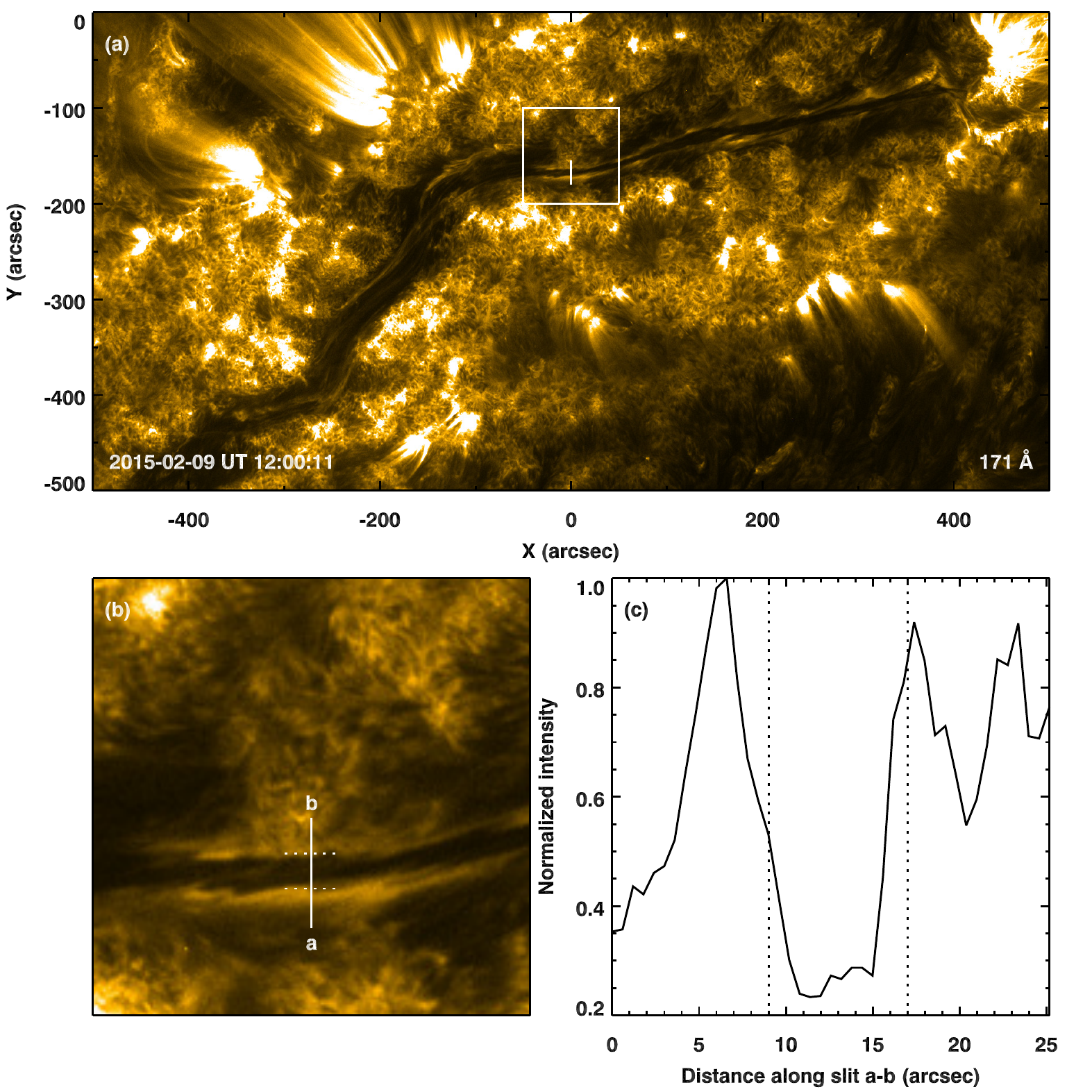}
\caption{Filament or the on-disk counterpart of the 2015 prominence. \textit{Panel a}: Same as Fig.\,\ref{fig:cont}a, but plotted for the observations obtained on 2015 February 9. The filament is visible as a dark elongated structure. The white box marks a section of the filament, identified by a solid white line, that crosses the central meridian. \textit{Panel b}: Zoom into the region marked by the white box in \textit{panel a}. Here the white line a-b has the same meaning as in \textit{panel a}. The shorter dotted lines (perpendicular to the solid line) cover the width of the filament. \textit{Panel c}: Intensity variation along slit a-b showing a dip at the location of the filament. The dotted vertical lines mark the width of the filament (same as in panel b), which is 8\arcsec (corresponding to about 6 Mm). See Appendix\,\ref{sec:evap}.
\label{fig:cont3}}
\end{center}
\end{figure*}

From visual inspection, we identified at least six events of propagating disturbances, each lasting $\sim$10 minutes, over a duration of 260 minutes in the 2015 case (Fig.\,\ref{fig:pd2}). Similarly, we identified four such events over a period of 120 minutes in the 2016 case (Fig.\,\ref{fig:pd3}). This suggests that at a given location, prominence spicules that give rise to propagating disturbances persist for 20\% to 30\% of time. Conservatively, we assume that their filling fraction is between 0.01 and 0.1 at any given time, meaning an area coverage of 1\% to 10\%. The density contrast between the prominence and corona, $\rho_\text{pro}/\rho_\text{cor}$, is about 100 \citep{2010SSRv..151..243L}. From Fig.\,\ref{fig:pd1}, we obtained $v=40$\,km\,s$^{-1}$ (the propagation speed of the intensity disturbance). For this example, we measure a filament width of about 10\,Mm in AIA 171\,\AA\ images (when the prominence crossed the central meridian on the solar disk on 2015 February 9, UT 12:00; Fig.\,\ref{fig:cont3}).

By substituting these values for the filling factor, density contrast, width, and velocity, we obtain timescales in the range of three days to four weeks. This evaporation timescale is much shorter than the lifetime of the prominence (about 20 weeks; Appendix\,\ref{sec:prom1}). In the absence of other processes that might replenish the filament, the prominence spicules evaporate it on timescales of days. This identifies a new channel for the decay of quiescent solar filaments.

\end{appendix}

\end{document}